\documentclass[final,1p,times,twocolumn]{elsarticle}

\usepackage{amssymb}
\usepackage{amsmath}
\usepackage{multicol}
\usepackage{float}
\usepackage{caption}
\usepackage{cleveref}
\usepackage{titlesec}

\crefname{figure}{Fig.}{Figs.}
\Crefname{figure}{Fig.}{Figs.}

\journal{Nuclear Physics B}

\titlespacing*{\section}
{0pt}{2pt}{2pt} 

\begin{document}

\begin{frontmatter}

\def\pmbanner{{\hrule height 1 pt}\vskip35pt{NIMA POST-PROCESS BANNER TO BE REMOVED AFTER FINAL ACCEPTANCE}\vskip35pt{\hrule height 4pt}\vskip20pt}


\title{From Simulation to Fabrication: Realizing Silicon Crystalline Undulators with Silicon Nitride Stressor Layer Patterning}


\author[infnferrara]{L. Malagutti\corref{cor1}}
\cortext[cor1]{First author, Corresponding author}
\ead{lorenzo.malagutti@fe.infn.it}

\author[infnferrara]{L. Bandiera}
\author[CNR]{F. Bonafè}
\author[infnferrara]{N. Canale}
\author[padova,legnaro]{D. De Salvador}
\author[ferrara]{P. Fedeli}
\author[infnferrara,ferrara]{V. Guidi}
\author[MBN]{A. V. Korol}
\author[CNR]{F. Mancarella}

\author[infnferrara,ferrara]{R. Negrello}
\author[infnferrara]{Gianfranco Paternò}
\author[infnferrara]{M. Romagnoni}
\author[padova,legnaro]{F. Sgarbossa}
\author[MBN]{A.V. Solov'yov}
\author[infnferrara]{A. Sytov}
\author[padova,legnaro]{D. Valzani}

\author[ferrara]{A.Mazzolari\corref{cor2}}
\cortext[cor2]{Corresponding author}
\ead{andrea.mazzolari@unife.it}

\address[infnferrara]{INFN, Sezione di Ferrara, Ferrara, Italy}
\address[ferrara]{Università degli Studi di Ferrara, Ferrara, Italy}
\address[padova]{Università degli Studi di Padova, Padova, Italy}
\address[legnaro]{INFN, Laboratori Nazionali di Legnaro, Legnaro, Italy}
\address[CNR]{CNR - IMM, Bologna, Italy}
\address[MBN]{MBN Research Center, Frankfurt, Germany }


\begin{abstract}

A crystalline undulator is a crystal exhibiting periodic deformations that cause channeled particles to oscillate, generating coherent electromagnetic waves. In this study, stressor layer patterning has been employed to induce the desired deformations on a silicon substrate. Finite element method simulations were performed to optimize the geometric parameters of the undulator. The primary focus was on achieving sinusoidal deformation with sub-millimeter period and amplitude exceeding 1~nm, utilizing the silicon (110) plane for its superior channeling properties. Key parameters, such as substrate thickness and undulator period, were meticulously refined to ensure uniform deformation while minimizing the impact of higher harmonics, which could degrade performance. Based on the simulation results, a crystalline undulator, 160 \( \, \mu\text{m}\) thick and consisting of 10 periods with a period length of 334 \( \, \mu\text{m}\), has been successfully fabricated. The final device demonstrates structural integrity and a uniform deformation extending up to 20 \( \, \mu\text{m}\) from the surface. Specifically designed for use with 5-30 \,\text{GeV} particle beams, the undulator is capable of generating gamma photons in the 5-15 \,\text{MeV} range. This work effectively integrates advanced simulation techniques with precise fabrication methods, demonstrating the feasibility of crystalline undulators as high-performance devices for generating gamma radiation.

\end{abstract}



\end{frontmatter}


\begin{multicols}{2}
\section{INTRODUCTION}
\label{sec:Intro}

The generation of high-intensity, monochromatic X-ray sources plays a pivotal role in driving advancements across a wide spectrum of scientific disciplines, from basic research to applied technologies. Although free-electron lasers (FELs) are widely utilized for producing intense soft X-ray beams, the generation of hard X-rays up to a few tens of keV is limited by the performance of magnetic undulators, which generate magnetic fields up to 10 tesla and have a minimum oscillation period of few centimeters \cite{Olive2014}. 
To overcome these limitations, the idea of a crystalline undulator was proposed in the 1980s \cite{baryshevsky-1980,kaplin1980charged}. The feasibility of this scheme was demonstrated in Refs. \cite{ikezi-1984,bogacz-1986,baryshevsky-1991,dedkov-1994, avakyan-1998,Korol1999,korol-2000,korol-2004}, and further developed in Refs. \cite{Baryshevsky2013_Important,Korol2014,kostyuk-2013}. 
The practical realization of CUs is the subject of the currently running Horizon Europe EIC-Pathfinder-2021 project TECHNO-CLS \cite{mbnresearch}. Crystalline undulator (CU) presents an equivalent magnetic field of thousands of tesla and could provide a period \( \lambda_{\text{CU}} \) in the sub-millimeter range, radiating a high flux of photons in the hard X and soft \(\gamma\)-ray range. The availability of a \(\gamma\) -ray source will enable the advancement of innovative applications in nuclear medicine \cite{Paterno2015}, industry \cite{Kraemer2018} and nuclear waste management \cite{Rehman2017}. 
In a CU, electrons or positrons are driven into oscillatory motion by the strong electrostatic field generated by the alignment of atoms along crystallographic planes or axes. Thanks to the channeling phenomenon \cite{lindhard-1964}, particles entering the crystal at a small angle relative to the primary crystallographic planes or axes are confined along specific trajectories.  
The channeling condition is satisfied when the particle's angle of incidence is smaller than the critical angle for channeling $\theta_c$, given by Eq. \ref{eq:ThetaC}:

\begin{equation}
 \theta_c = \sqrt{\frac{2 U_0}{p v}}    ,
\label{eq:ThetaC}
\end{equation}

Where $U_0$ is the depth of the potential well, $p v$ the particle's relativistic energy, $p$ the momentum and $v$ the velocity of the particle.

The oscillatory motion of channeled particles is accompanied by the emission of channeling radiation \cite{Kumakhov1976}. However, in this scenario, the oscillations are not in phase, preventing the system from achieving an undulator-like regime. In contrast, a CU presents a periodic bending deformation of the crystal lattice, which forces the particle to oscillate with the same frequency, generating intense radiation peaks \cite{Korol2014}. The energy of a photon, E, emitted in an undulator is proportional to the square of the particle Lorentz factor \(\gamma\) and inversely proportional to the undulator period \( \lambda_{\text{CU}} \) following Eq. \ref{eq:Energy} :

\begin{equation}
 E = \frac{2\hbar c{\gamma^2}}{\lambda_{\text{CU}} \left( 1 + \frac{K^2}{2} \right)}      ,
\label{eq:Energy}
\end{equation}

where \(\hbar\) is Planck's constant, \(c\) is the speed of light and \(K\) is the deflection parameter. The parameter \(K\) is defined by Eq. \ref{eq:K_Eq} :

\begin{equation}
K = 2\pi \gamma \frac{A_{\text{CU}}}{\lambda_{\text{CU}}}       ,
\label{eq:K_Eq}
\end{equation}

where \(A_{\text{CU}}\) is the amplitude of the transverse oscillations of the particle beam, \(\gamma\) is the Lorentz factor, and \(\lambda_{\text{CU}}\) is the undulator period.
The capability of the CU lies in its precise tunability, allowing adjustments to crystal parameters to achieve specific energy peaks. \\[-\baselineskip]

A crystal, typically made of materials such as Si, Ge, or diamond, can withstand substantial stress while remaining within the elastic regime. A complex geometry, such as the periodically bent crystal needed to drive the particles inside manipulated channels, can be achieved. In order to achieve the periodically bent deformation, different techniques have been developed and evaluated through x-ray analysis \cite{Bellucci2003,Bellucci2003_2,Camattari2019}, such as mechanical scratching \cite{Bellucci2003}, the laser ablation technique \cite{Balling2009}, the grooving method \cite{Guidi2005,Bagli2014}, graded composition strained layers in superlattices \cite{Mikkelsen2000,Avakian2002}, tensile/compressive strips deposition \cite{Guidi2007,Guidi2011}, sandblasting \cite{Camattari2017} ion implantation \cite{Bellucci2015} and ultrasonic waves \cite{Korol1999,Avakian2001}. \\[-\baselineskip]

The aim of the present work is to design a silicon CU via the finite element method (FEM) and its fabrication exploiting a tensile silicon nitride (\( \text{Si}_3\text{N}_4 \)) stressor layer patterning in order to obtain a usable real undulator. 
The use of silicon and silicon nitride as a stressor layer is a widely adopted method in microelectronics, offering fewer defects, more uniform deformation, and greater process standardization compared to other fabrication techniques. Additionally, it ensures long-term durability, stability, and reliability, making the process reproducible, scalable, and well-suited for the development of high-quality undulators. 
\section{Simulation via Finite Element Method}
\label{sec:Simulation}

\begin{figure}[H]
    \centering
    \includegraphics[width=\linewidth]{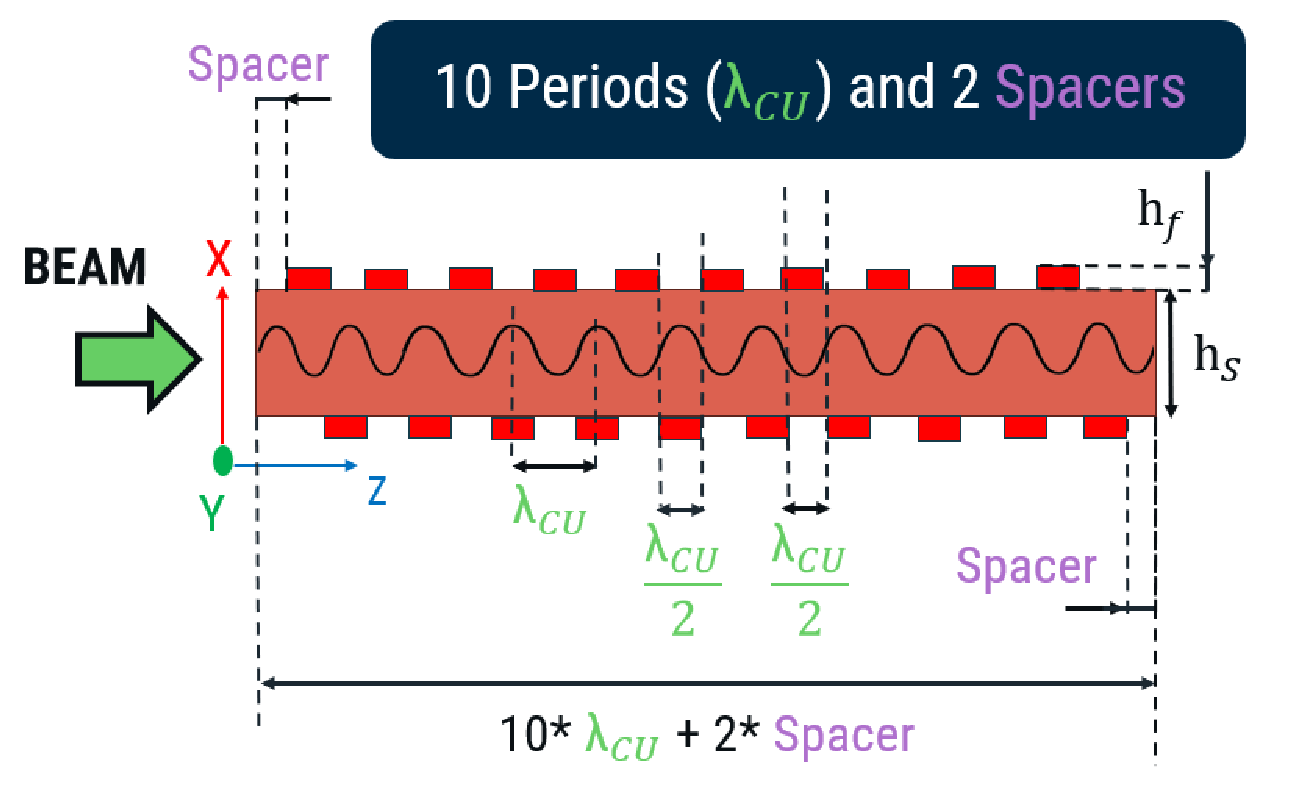}
    \caption{ Cross section of the undulator along the beam shows the layer patterning of nitride stressor and main geometric parameters: thickness of substrate \( h_{\text{s}} \) and film \( h_{\text{f}} \) , undulator period \( \lambda_{\text{CU}} \), number of periods 
    and spacer.}
    \label{fig:cross-section}
\end{figure}

In this section, we outline the key parameters and describe the simulation setup used to assess the sinusoidal deformation of the CU, focusing on large amplitudes in the hundreds of micrometers period range \cite{Korol2022}. Several conditions must be fulfilled to obtain a functioning CU \cite{Korol2014}, the most relevant from a geometric perspective being that amplitude of the sinusoidal deformation \(A_{\text{CU}} \) must be much greater than interplanar distance \( d_p\), corresponding to the large-amplitude regime. Considering a silicon (1 1 0) plane with an interplanar distance \( d_p = 1.92 \)  \(\text{Å} \), the \( A_\text{CU} \) should exceed \( 1 \, \text{nm} \).
Another important condition is that the number of periods N, must be significantly greater than one to ensure the large-amplitude regime. Therefore, the number of periods has been set to 10 to observe undulation regime in confined thickness across the beam. In order to evaluate the best combination of geometric parameters to produce a real undulator suitable for 5-30 \text{GeV}, the finite element method  has been employed through Ansys software \cite{ansys}. The behavior of a silicon CU with a sub-millimeter period obtained with silicon nitride stressor layer patterning as reported in \cref{fig:cross-section} can be estimated. The thickness of the substrate and the film are respectively \( h_{\text{s}} \) and \( h_{\text{f}} \). A pattern of \( \text{Si}_3\text{N}_4 \) rectangular strips with a period equal to \( \lambda_{\text{CU}} \) along the beam direction are present on both sides shifted by \( \frac{\lambda_{\text{CU}}}{2} \) (\cref{fig:cross-section}). To reach the maximum amplitude of the sinusoidal deformation induced by the stressor strips of \( \text{Si}_3\text{N}_4 \), a strip width  equal to \( \frac{\lambda_{\text{CU}}}{2} \) \cite{Lanzoni2008} has been chosen in the simulation. 

A wafer coated on one side by an \( \text{Si}_3\text{N}_4 \) film presents a radius of curvature R and a film stress \( \sigma_{\text{f}} \) completely in agreement with Stoney law (Eq. \ref{eq:Stoney}) \cite{Stoney1909,Guidi2007, Lanzoni2008}:

\begin{equation}
 \sigma_f = \frac{E_s h_s^2}{6 R h_f (1-\nu_s)}          ,
\label{eq:Stoney}
\end{equation}

where \( h_{\text{s}} \)  , \( E_{\text{s}} \)  , \( \nu_{\text{s}} \)   are, respectively, the thickness, Young's modulus and Poisson's ratio of the substrate and \( h_{\text{f}} \) is the thickness of the film.

The stress \( \sigma_{\text{f}} \) in a \( \text{Si}_3\text{N}_4 \) film, deposited by low pressure chemical vapor deposition (LPCVD) on a Si substrate is the result of thermal and intrinsic stresses \cite{Thurn2002}. Since LPCVD takes place at temperatures typically around 800 °C, thermal stress arises when the wafer is cooled, due to the difference in thermal expansion coefficients between the the substrate \( \alpha_s \) and the film \( \alpha_f \). From a simulation point of view, one approach to simulate the residual effects caused by the film deposition process is to subject the silicon substrate to an equivalent thermal load, \(\Delta T\), that generates the same stress in the coating following the Eq. \ref{eq:deltaT}: 

\begin{equation}
 \Delta T \approx \frac{\sigma_f}{\left( \frac{E_f}{1 - \nu_f} \right) (\alpha_s - \alpha_f)}         ,
\label{eq:deltaT}
\end{equation}

Taking into account the (110) plane, the simulations have been performed assuming the anisotropic mechanical properties of silicon \cite{Lanzoni2008} and isotropic thermal properties. For \( \text{Si}_3\text{N}_4 \), both the mechanical and thermal properties have been considered isotropic. Except for the mechanical properties of silicon \cite{Lanzoni2008}, the material properties are: \( \alpha_s = 3 \times 10^{-6} \)  \(\, \text{C}^{-1} \) , Young's modulus of film, \( E_f \) =  310  \(\,GPa\) , poisson's coefficient of film, \( \nu_f \) = 0.24 and \( \alpha_f = 1.1 \times 10^{-6} \, \ \text{C}^{-1} \) \cite{Lanzoni2008}. In addition, as an input parameter for the simulations, the thickness of the film \(h_f\) and \( \sigma_{\text{f}} \) have been set constant respectively equal to 400 \(nm\) and 800 \( MPa\)  as these are the standard in the microelectronic. Using formula in Eq.\ref{eq:deltaT}, a \( \Delta T\)  \(\approx 1030 \) \(\,^\circ\text{C} \) must be applied to induce a stress of \(800 \, MPa\) in the substrate. All mechanical and geometric parameters have been defined, with the exception of  substrate thickness \( h_{\text{s}} \) and undulator period \(\lambda_{\text{CU}} \). These two parameters have been varied during simulations to evaluate the optimal combination that can be feasibly manufactured and tested experimentally. Several simulations were carried out to analyze how these parameters influence the sinusoidal deformation at different heights along the substrate thickness and parallel to the beam direction. \\[-\baselineskip]

To ensure proper functioning of the undulator, the deformation must exhibit an amplitude greater than \(1nm\) and closely approximate a sinusoidal shape, which corresponds to having the lowest possible number of harmonics in the Fourier spectrum. The Fourier spectrum of the sinusoidal directional deformation along the thickness shows an increasing number of harmonics moving from the center to the outer layer of the CU resulting in a quasi-sinusoidal deformation for three-fourths of the crystal \cite{Guidi2007}. The spectrum can be decomposed in two main terms, the first harmonic and the total contribution of all the other harmonics. The sinusoidal deformation amplitude generated from the first harmonic is proportional to \(\left(\frac{h_s}{\lambda_{\text{CU}}}\right)^2\) and for \(\frac{h_s}{\lambda_{\text{CU}}}\) greater than 2 the amplitude decreases moving outward. The ratio \(\frac{h_s}{\lambda_{\text{CU}}}\) is also crucial in determining the total contribution of the other harmonics to the deformation and for values greater than 2.5, this contribution remains quite uniform at \(\approx 10\% \) of the first harmonic. \\[-\baselineskip]

\begin{figure}[H]
    \centering
    \includegraphics[width=\linewidth]{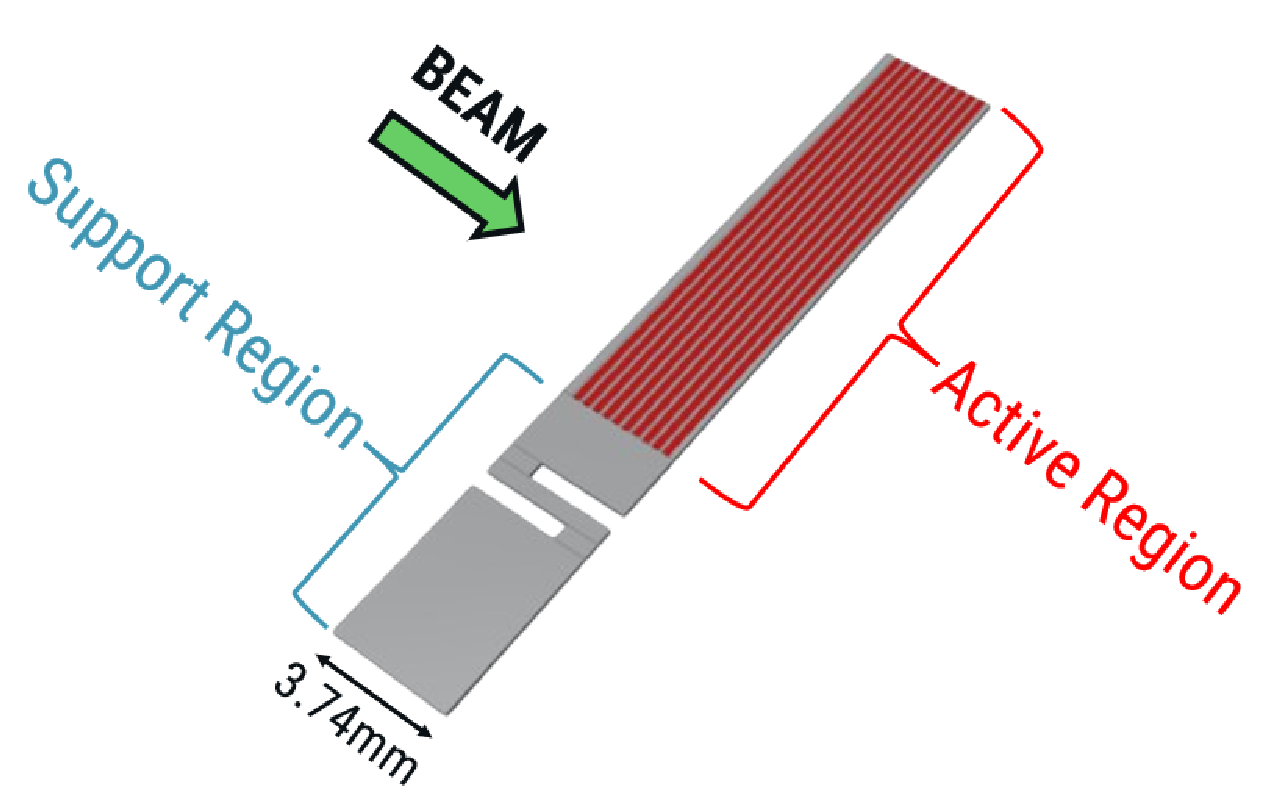}
    \vspace{0.05cm}
    \caption{Real geometry of crystalline undulator with a width along the beam of 3.74mm and divided in two zones: support and active region.}
    \label{fig:geometry real}
\end{figure}

From a fabrication perspective, a lower limit of \(h_s = 160 \) \(\, \mu\text{m}\) was chosen to ensure proper handling and structural integrity of the undulator. To reduce the entrance angle below the critical channeling angle \(\theta_c\), which for a 10 GeV beam is approximately \(70 \, \mu\text{rad}\), the spacer length (\cref{fig:cross-section}) was set to 200 \( \, \mu\text{m}\) ensuring that the incidence angle remains well below \(\theta_c\), thereby maintaining channeling efficiency. The final geometry of the undulator is shown in \cref{fig:geometry real} and consists of two zones: a supported area and an active area where the beam passes, measuring \(10 \) and \(20 \, \text{mm}\) in length, respectively. A stress relief cut is present in the supported area to prevent any deformation in the active area due to clamping during the experimental test. This ensures that the clamping forces do not affect the active area or distort its deformation.

\captionsetup[figure]{skip=-10pt} 

\begin{figure}[H]
    \centering
    \includegraphics[width=\linewidth]{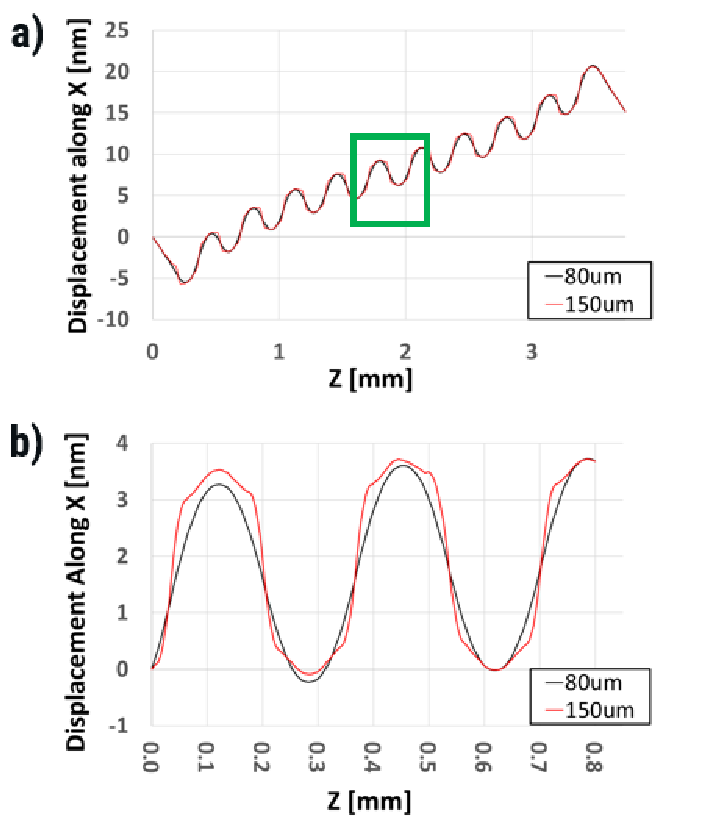}
    \vspace{0.12cm}
    \caption{a) shows the total deformation of the undulator along the beam at two different height in the middle of active area: at the center and 10 \(\mu\text{m}\) from the surface of the undulator;   b) displays the sinusoidal deformation of the green rectangle presents in a).}
    \label{fig:sinusoidal deformation}
\end{figure}

The actual geometry (\cref{fig:geometry real}) has been simulated with a period \(\lambda_{\text{CU}} = 334 \, \, \mu\text{m}\), as it provides the best combination for achieving both the most uniform deformation along the substrate thickness and a deformation amplitude significantly greater than \(1nm\). The final undulator has a width of 3.74 \( \, \text{mm}\), determined by the presence of 10 periods (\(\lambda_{\text{CU}} = 334\) \( \mu\text{m}\)) and two spacers of \(200 \,\, \mu\text{m}\). The sinusoidal deformation is reported in \cref{fig:sinusoidal deformation} at two different heights within the middle of the active area: at the center and 10 \(\mu\text{m}\) from the surface of the undulator. From \cref{fig:sinusoidal deformation}.b is clearly visible that the number of harmonics increases, and the deformation becomes progressively less sinusoidal as it approaches the surface. The oscillation amplitude is greater than \(1 \, \text{nm}\), specifically around \(1.6 \, \text{nm}\).

\captionsetup[figure]{skip=-7pt} 

\section{Fabrication}
\label{sec:Fabrication}

\begin{figure}[H]
    \centering
    \includegraphics[width=\linewidth]{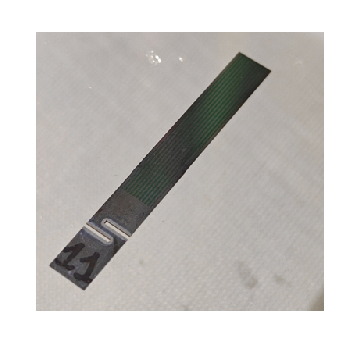}
    \caption{ Real manufactured crystalline silicon undulator.}
    \label{fig:real undulator}
\end{figure}

After performing the simulations and defining the parameters for constructing the crystalline undulator, a 160~\textmu m thick wafer with (110) plane, zero dislocations, and double-coated with a \(400 \,\, \text{nm}\) \( \text{Si}_3\text{N}_4 \) stressor layer has been purchased for processing. The removal of \( \text{Si}_3\text{N}_4 \) has been carried out on both sides of the wafer, with each side having a pattern shifted by \( \frac{\lambda_{\text{CU}}}{2} \) (\cref{fig:cross-section}). 
To achieve this, two different photolithographic masks, named front and back, have been designed to create an undulator with 10 periods of \(334 \,\, \mu\text{m}\) . Depending on the chosen parameters, multiple undulators can be obtained from a single wafer. The silicon wafer has been dehydrated in an oven at \( 110 \,^\circ\text{C} \) C for 30 minutes and subsequently exposed to hexamethyldisilazane vapor at \( 90 \,^\circ\text{C} \) to improve the adhesion of the photoresist. A 1~\textmu m thick layer of positive photoresist (Fujifilm HPR 504) has been deposited by spinning at 3000~rpm, followed by a thermal treatment at \( 110 \, ^\circ\text{C} \) for 60~seconds. The photoresist has been exposed to Near-UV light for 5~seconds using a SuSS MA4 aligner exploiting the front mask, then developed with AZ 726 MIF solution, rinsed with deionized water, and dried in a spin dryer using hot nitrogen. It has been further hard-baked on a hotplate at \( 90 \, ^\circ\text{C} \) for 60~seconds. Subsequently, 400~nm of \( \text{Si}_3\text{N}_4 \) have been removed using Reactive Ion Etching (RIE) with fluorine-based plasma (CHF\textsubscript{3} and Argon) for 90~minutes. The photoresist has been stripped using a microwave oxygen plasma asher. This process has been repeated for the other side of the wafer using the back photolithography mask. Using a Disco DAD 3320 dicing machine, multiple undulators, each 30~mm long, have been obtained with a spacer of  \(200 \,\, \mu\text{m}\) at the entrance and exit of the undulator.
Alignment of the \( \text{Si}_3\text{N}_4 \) strips on the front and back sides of individual undulators has been assessed using the Fogale Nanotech Tmap Dual IR300M, which has revealed an alignment precision of approximately 3~\textmu m, with the best alignment observed in the undulators located at the center of the wafer. Achieving such precision is challenging due to the alignment of the second mask after substrate deformation caused by \( \text{Si}_3\text{N}_4 \) removal from one side, especially given the wafer's thinness.
A strain relief cut has been performed on the crystalline undulator using a Keyence MD-U 1000C UV laser, as shown in \cref{fig:real undulator}.

This meticulous process has led to the successful fabrication of a crystalline undulator with precise alignment and structural integrity, demonstrating the feasibility of producing such devices with high precision on thin silicon substrates for advanced applications.

\section{Conclusion}
\label{sec:Conclusion}

This study presents a comprehensive design and fabrication strategy for crystalline undulators. The FEM simulations determined the optimal parameters to produce a crystal undulator with a 334 micron period, ideally suited for positron beams of 5–30 \text{GeV}. A reliable fabrication process based on photolithographic fabrication process has been described and validated by precise infrared microscopy alignment, overcomes critical challenges such as pattern alignment on opposite crystal faces. 
This work provides a novel approach for the fabrication of large-amplitude crystal undulators, contributing to the advancement of technology needed to use them as powerful and efficient sources of hard X-rays and gamma-rays, unlocking new frontiers in scientific exploration and innovation.

\section{Acknowledgement}
\noindent This work was supported by the European Commission through the  MHz-TOMOSCOPY (G.A 101046448), H2020-MSCA-RISE N-LIGHT (G.A. 872196), EIC-PATHFINDER-OPEN TECHNO-CLS (G.A. 101046458), and H2020-MSCA-IF-Global TRILLION (G.A.101032975) projects.

We acknowledge the CNR, Institute for Microelectronics and Microsystems (IMM), Italy, for their scientific collaboration and valuable support in this work.

We acknowledge financial support under the National Recovery and Resilience Plan (NRRP), Call for tender No. 104 published on 02.02.2022 by the Italian Ministry of University and Research (MUR), funded by the European Union – NextGenerationEU – Project Title : "Intense positron source Based On Oriented crySTals - e+BOOST" 2022Y87K7X – CUP I53D23001510006.

 \bibliographystyle{elsarticle-num} 
 \bibliography{cas-refs}




\end {multicols}{2}
\end{document}